\documentclass[3p,times,twocolumn]{elsarticle}

\usepackage{ecrc}

\volume{00}

\firstpage{1}

\journalname{Nuclear and Particle Physics Proceedings}

\runauth{D. Caprioli}

\jid{nppp}

\jnltitlelogo{Nuclear and Particle Physics Proceedings}

\usepackage{amssymb}
\usepackage{hyperref}
\usepackage{color} 

\usepackage[figuresright]{rotating}

\newcommand{\Mpc}{\,Mpc$^{-3}$}

\newcommand{\cmq}{\,cm$^{-2}$}
\newcommand{\yr}{\,yr$^{-1}$}
\newcommand{\s}{\,s$^{-1}$}
\newcommand{\ergs}{\mbox{\,erg\,\s}}
\renewcommand{\deg}{^{\circ}}
\newcommand\fdg{\mbox{$.\!\!^\circ$}}

\newcommand{\mui}{\mu_{\rm i}}
\newcommand{\muf}{\mu_{\rm f}}
\newcommand{\xii}{\xi_{\rm i}}

\newcommand{\rmin}{r_{\rm min}}

\begin{document}

\begin{frontmatter}


\author{Damiano Caprioli}
\ead{caprioli@uchicago.edu}
\ead[http://astro.uchicago.edu/~caprioli/]{http://astro.uchicago.edu/~caprioli/}
\address{University of Chicago, Department of Astronomy \& Astrophysics,
5640 S Ellis Ave., Chicago, IL 60637, USA}

\dochead{}

\title{An Original Mechanism for the Acceleration of Ultra-High-Energy Cosmic Rays}




\begin{abstract}
We suggest that ultra-high-energy (UHE) cosmic rays (CRs) may be accelerated in ultra-relativistic flows via a one-shot mechanism, the 
 ``espresso" acceleration, in which already-energetic particles are generally boosted by a factor of $\sim\Gamma^2$ in energy, where $\Gamma$ is the flow Lorentz factor. 
More precisely, we consider blazar-like jets with $\Gamma\gtrsim 30$ propagating into a halo of ``seed" CRs produced in supernova remnants, which can accelerate UHECRs up to $10^{20}$\,eV.
Such a re-acceleration process naturally accounts for the chemical composition measured by the Pierre Auger Collaboration, which resembles the one around and above the knee in the CR spectrum, and is consistent with the distribution of potential sources in the local universe; particularly intriguing is the coincidence of the powerful blazar Mrk 421 with the hotspot reported by the Telescope Array Collaboration.
\end{abstract}


\end{frontmatter}


\section{The Origin of Cosmic Rays} 
More than one hundred years after their discovery, the origin of the highest-energy extraterrestrial particles still remains one of the most challenging questions in physics. 
The new generation of telescopes and detectors seems to support the so-called supernova remnant (SNR) paradigm for the origin of CRs below $10^{17}$\,eV \cite[e.g.,][]{tycho}: nuclei can be accelerated in SNR blast waves via diffusive shock acceleration \citep[with efficiency of 10-20\%, see][]{DSA,MFA,diffusion,injection,AZ} in a rigidity-dependent way, so that the maximum energy scales with the ion charge $E_{\rm max}\propto Z$, possibly up to the CR ``knee'' if protons can achieve a few PeV in some young remnants.
The imprint of such a dependence on the ion charge is visible in CR data \cite[e.g.,][]{hoerandel+06,argo12}, which shows that the chemical composition becomes heavier above the knee as a result of a cut-off in rigidity in the PeV range \cite{nuclei,AZ}.

Acceleration of higher-energy CRs up to 10$^{20}$\,eV, instead, does not have such a prominent source candidate.
The astrophysical objects (potentially) able to generate Ultra-High-Energy CRs (UHECRs) comprise gamma-ray bursts (GRBs) \cite[e.g.,][]{waxman95,vietri95}, newly-born millisecond pulsars \cite[e.g.,][]{Blasi+00,Fang+12}, and active galactic nuclei (AGNs) \cite[e.g.,][]{ostrowsky00,Murase+12,espresso}.

The measurement of UHECR chemical composition is improving dramatically thanks to the increasing statistics and the better reconstruction of extensive atmospheric air showers induced by UHECRs.
The Pierre Auger Observatory recently reported a composition compatible with pure protons above $10^{17}$\,eV, but 
not in the ankle region: even below $10^{19}$\,eV nuclei heavier than helium are necessary to explain the observed correlation between the depth of shower maximum and the signal in the water Cherenkov stations of air-showers registered simultaneously by the fluorescence and the surface detectors \citep{Auger14a,Auger16}.
These findings are not inconsistent with Telescope Array (TA) data, which favor a lighter chemical composition on the whole energy range, if statistics, nuclear interaction models, and pipeline analysis are taken into account \citep{Pierog13, PAO-TA15}.
The lack of statistics above  $10^{19}$\,eV does not allow us to draw conclusions on the nature of the highest energy CRs, but data support the idea that the chemical composition becomes heavier and heavier with energy, possibly consistent with an iron-dominated population close to the  cut-off of the CR spectrum.

Here we summarize the features of a recently-proposed mechanism for the acceleration of UHECRs in AGN jets that only relies on basic properties of relativistic flows, the \emph{espresso} mechanism \cite[][hereafter, C15]{espresso}. 
After a brief review of such a model, in which ultra-relativistic AGN jets provide a \emph{one-shot} re-acceleration of Galactic-like CRs accelerated in SNRs, we discuss the obtained CR spectrum and chemical composition from the knee to the cut-off (\S\ref{sec:esp}) and the  possibility of tracing back UHECRs events to local sources.
In particular, we assess the consistency of the proposed scenario with the local AGN distribution, also pointing out the coincidence of the TA hotspot above $5.7\times 10^{19}$\,eV and the position of Mrk 421 (\S\ref{sec:sources}).
Finally, in \S\ref{sec:traj} we calculate the orbits of representative re-accelerated particles in simplified jet structures, showing that energy gains $\gtrsim \Gamma^2$ are almost always guaranteed regardless of the jet magnetic structure and initial particle momentum.

\section{The \emph{Espresso} Acceleration of UHECRs}\label{sec:esp}
The main idea behind the espresso acceleration is that energetic particles with gyroradii large enough to cross the boundary between a relativistic flow (with Lorentz factor $\Gamma$) and the ambient medium can experience a Compton-like scattering, gaining a factor of $\sim\Gamma^2$ in energy [C15].
Such a one-shot acceleration occurs if the ingoing and outgoing flight directions are uncorrelated \cite{Achterberg+01}, equivalent to the very mild requirement that particles spend at least a fourth of a gyro-period in the flow before being released. 

More precisely, we consider Galactic-like CRs accelerated in SNRs up to $10^{17}$\,eV \cite[e.g.,][]{nuclei} as seeds to be re-accelerated in powerful AGN jets with Lorentz factors as large as $\Gamma\simeq 30$ or more \cite[e.g.,][]{Tavecchio+10}. 
A boost of order $\sim\Gamma^2\approx 10^3$ would therefore be sufficient to produce UHECRs with energies up to $10^{20}$\,eV.

Such a re-acceleration model predicts that the chemical composition of CR seeds, increasingly heavy above $10^{13}$\,eV and iron-like around $10^{17}$\,eV \cite[e.g.,][]{KU12}, should be reflected into that of UHECRs. 
This scenario is supported by Auger data, which is consistent with a proton-only flux at $10^{18}$\,eV and a nitrogen-like composition at $\sim 4\times 10^{19}$\,eV \citep{Auger14a, Auger16}.
The fact that the UHECR composition becomes heavier and heavier with energy suggests a possible iron contribution at the highest energies, where the limited statistics does not allow conclusive claims, yet.

\begin{figure}[t]
\begin{center}
\includegraphics[trim=30px 40px 30px 40px, clip=true, width=.5\textwidth]{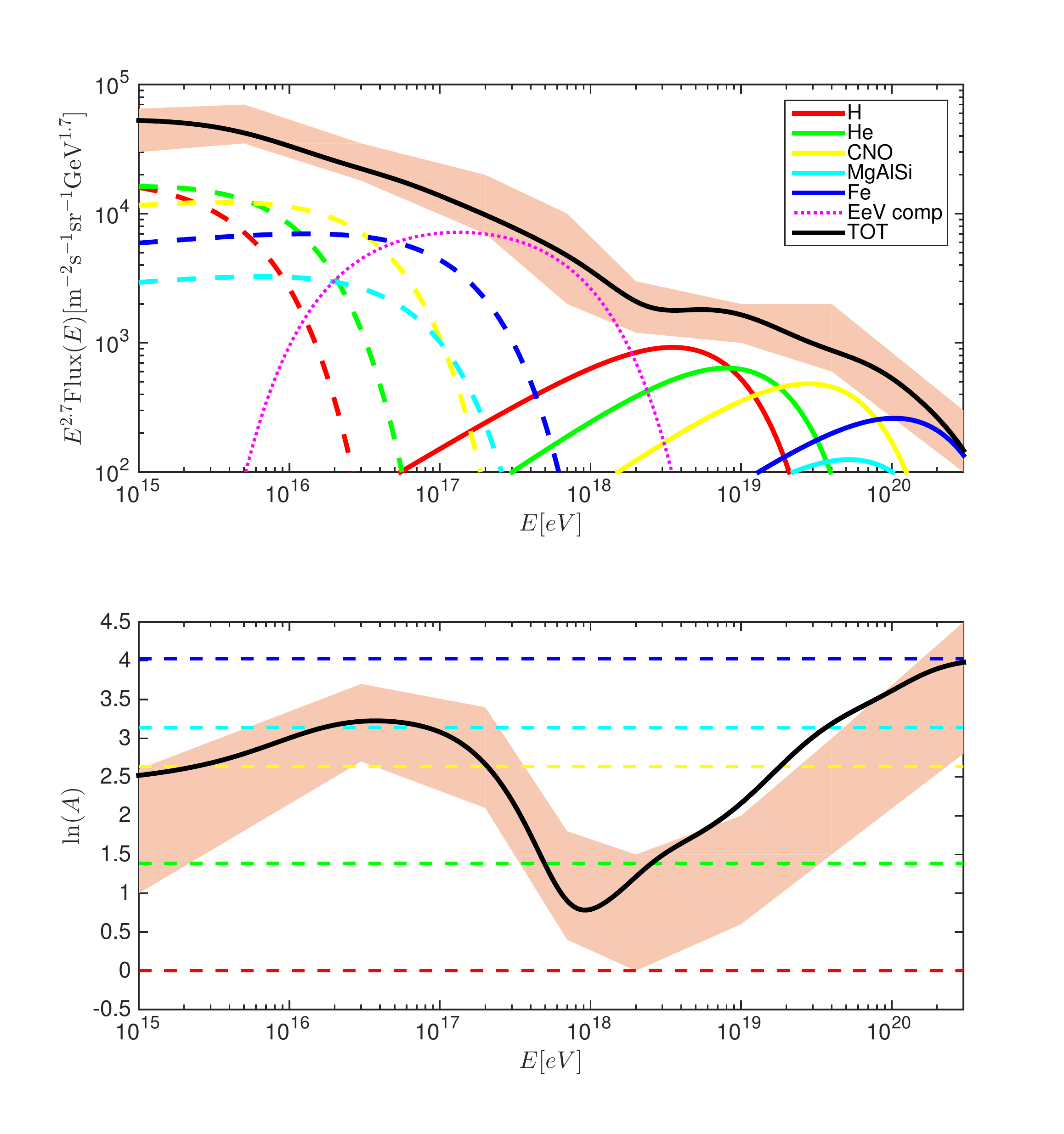}
\caption{Top panel: fluxes of different CR species above $10^{15}$\,eV, as in the legend. 
Dashed and solid lines correspond to CRs produced in SNRs and to UHECRs espresso-accelerated in AGN jets with $\Gamma\simeq 30$.
Bottom panel: predicted average atomic mass as a function of energy; colors correspond to the elements of the top panel.
The bands illustrate a convolution of data from various experiments \citep[][and refs.\ therein]{KU12,GST13,Auger14a,Auger14b}.
See C15 for more details.
}
\label{fig:spec}
\end{center}
\end{figure}

Fig.\ \ref{fig:spec} (from C15) shows fluxes and chemical composition above $10^{15}$\,eV for a population of Galactic-like CRs re-accelerated in a jet with $\Gamma\simeq 30$.
The model qualitatively reproduces the characteristic spectral features and the observed light--heavy--light--heavy modulation in the CR composition.
The relative normalization of the two components is chosen in order to match observations.
More details about the choice of parameters that lead to fig.\ \ref{fig:spec} as well as discussions on the universality of the knee feature in other galaxies and the efficiency of re-acceleration can be found in C15.

Modeling the whole CR spectrum requires to account for the contributions of (at least) two classes of sources, typically one Galactic and one extra-galactic;
a single-source origin for all of the CRs seems unlikely given the non-monotonic chemical composition as a function of energy.
The observed spectrum also depends on the CR production/confinement volume, which varies from Galactic scales below $10^{17}$\,eV, to the whole universe up to $\sim 10^{19}$\,eV, to the local universe at the highest energies. 
Yet, the spectrum manages to be quite smooth, without any abrupt jump: this apparent fine tuning remains an unsolved question.
Where the espresso mechanism succeeds in removing ad-hoc prescriptions is in predicting the observed chemical composition, a feature shared by any other re-acceleration scenario (even if ---to our knowledge--- this general property has never been put forward before).

In addition to the normalization, CR models must reckon with explaining the observed power law indexes, too. 
Models of the recent UHECR data favor rather flat spectra $\propto E^{-1}$--$E^{-1.5}$ \cite[][]{GST13,abb14,Taylor14}, significantly harder than the spectrum of Galactic CRs;
no universal explanation has been proposed for such flat spectra in any acceleration scenario.
At the zero-th order, the espresso mechanism may be thought to replicate the slope of the knee region $(\sim E^{-3}$) at higher energies, but there are several  processes that may alter this expectation, for instance:
i) at high galactic latitudes the CRs reprocessed by AGN jets are expected to exhibit a flatter spectrum, especially if spallation losses dominate over diffusive escape [C15];
ii) seed acceleration at the jet boundary layer \cite{ostrowsky00}, energy-dependent penetration into the jet, and/or a jet with stratified Lorentz factors may systematically favor the re-acceleration of higher-rigidity particles;
iii) depending on the value of the intergalactic magnetic field, the extra-galactic spectrum may have a low-energy cutoff that prevents low-energy UHECRs from reaching the Earth in less than a Hubble time \cite[e.g.,][]{mr13,as14}.

\section{Constraints on Potential UHECR Sources}\label{sec:sources}
In order to accelerate particles up to $10^{20}$\,eV, an accelerator must satisfy three constraints: energetics, luminosity, and confinement.
AGNs easily meet these requirements, if the highest-energy particles are heavy nuclei \citep[e.g.,][]{espresso,murase09}.
Moreover, UHECR sources must be distributed in the universe in such a way that attenuation processes do not prevent the highest energy particles to reach the Earth, i.e, some sources must reside within the UHECR horizon.

The energy generation rate necessary to account for the UHECR flux above $10^{18}$\,eV is \citep[e.g.,][]{Katz+13}
\begin{equation}\label{eq:energy}
\dot\varepsilon_{\rm >EeV} \sim 5.4\times 10^{45}\,{\rm erg} $\Mpc \yr$.
\end{equation}
Assuming that the density of radio-loud active galaxies is $\approx10^{-5} $\Mpc  and that the AGN bolometric luminosity lies in the range $10^{43}\ergs\lesssim L_{\rm bol}\lesssim 10^{48}\ergs$ \citep{Jiang+07,woo-urry02}, one gets
$\dot\varepsilon_{\rm >EeV} \approx 10^{-2} \dot\varepsilon_{\rm AGN,bol}$.
Moreover, note that the actual AGN jet power (be it magnetically or kinetically dominated), which fuels both photons and energetic particles, is inferred to be at least one/two orders of magnitude larger than $L_{\rm bol}$ \cite[e.g.,][]{GTG09}.

The acceleration up to an energy $E_{\rm max}$ within a dynamical timescale (e.g., the expansion rate of a relativistic flow) requires a minimum source luminosity \citep[][]{Waxman04}:
\begin{equation}\label{eq:Lcr}
L_{\rm min}\approx 5\times 10^{42}~\frac{\Gamma^2}{\beta}\left(\frac{E_{\rm max}}{Z_{26}\, 10^{20}{\rm \, eV}}\right)^2 \mbox{erg\s},
\end{equation}
where $\Gamma$ and $\beta c$ are the flow Lorentz factor and speed, and $Z_{26}$ is the CR charge in units of that of a fully-ionized iron nucleus.
The jet luminosity of powerful AGNs can easily exceed $L_{\rm min}$, provided that UHECRs with $E_{\rm max}\sim 10^{20}$\,eV are heavy nuclei and that $\Gamma\lesssim 50$.

Finally, the Hillas criterion \citep{hillas84} sets the minimum requirements on source magnetic field and size for acceleration up to $E_{\rm max}$, and reads
\begin{equation}\label{eq:hillas}
B_{\mu{\rm G}}R_{\rm kpc}\gtrsim \frac{4\beta}{Z_{26}}\frac{E_{\rm max}}{10^{20}{\rm eV}},
\end{equation}
where $B$ is measured in $\mu$G and the source size $R$ in kpc.
Though often considered a confinement criterion (requiring CR gyroradius $\lesssim$ than $R$), this constraint in reality expresses the concept that acceleration up to $E_{\rm max}$ requires an induction electric field ($\sim\beta B$) to extend over a length $R$.
AGN jets satisfy the Hillas criterion because they are several kpc long and inferred to have fields of several to hundred $\mu$G.

\begin{table*}[htp]
\caption{Position, redshift, luminosity distance, and TeV emission of known blazars within 200 Mpc. From \href{http://tevcat.uchicago.edu}{\color{blue}{\tt http://tevcat.uchicago.edu}}.}
\begin{center}
\begin{tabular}{|l|c|c|c|c|c|c|}
\hline
Name 			& R.A. 				& DEC 		& z			& $d_L$ (Mpc)	& TeV Flux (\% Crab)	& TeV Luminosity (Crab)\\
\hline
IC310			& 49\fdg 18			& +41\fdg 32	& 0189	& 81			& 2.5				& $4.1\times 10^7$\\
Mrk421 			& 166\fdg 11		& +38\fdg 19	& 0.031		& 134.1			& 30				& $1.35\times 10^9$\\
Mrk501	 		& 253\fdg 48		& +39\fdg 76	& 0.034		& 147.4			& $>40$				& $>2.17\times 10^9$\\
1ES 2344+514 	& 356\fdg 77		& +51\fdg 70	& 0.044		& 192.2			& 11				& $1.02\times 10^9$\\
Mrk180	 		& 174\fdg 11		& +70\fdg 16	& 0.045 	& 196.7			& 11				& $1.06\times 10^9$\\
1ES 1959+650 	& 300\fdg 00		& +65\fdg 15	& 0.048		& 210.3			& 64				& $6.48\times 10^9$\\
AP Lib			& 229\fdg 43		& -24\fdg 37	& 0.049		& 214.9			& 2					& $2.31\times 10^8$\\
\hline
\end{tabular}
\end{center}
\label{tab:blazar}
\end{table*}

A straightforward implication of the observed heavy chemical composition is that the UHECR horizon is not set by photopion production on the cosmic microwave background radiation as suggested by Greisen, Zatsepin, and Kuz'min \cite{greisen66,ZK66}, but rather by photo-disintegration \cite[see, e.g.,][]{abb14,Dermer07,ko11}.
Surprisingly enough, the two radically attenuation processes make the universe transparent to UHECRs below $10^{19}$eV and return a comparable loss length at the highest energies, $\sim 100$\,Mpc at $10^{20}$eV for both proton and iron nuclei.

\begin{figure}[htbp]
\begin{center}
\includegraphics[trim=0px 0px 1px 0px, clip=true, width=.5\textwidth]{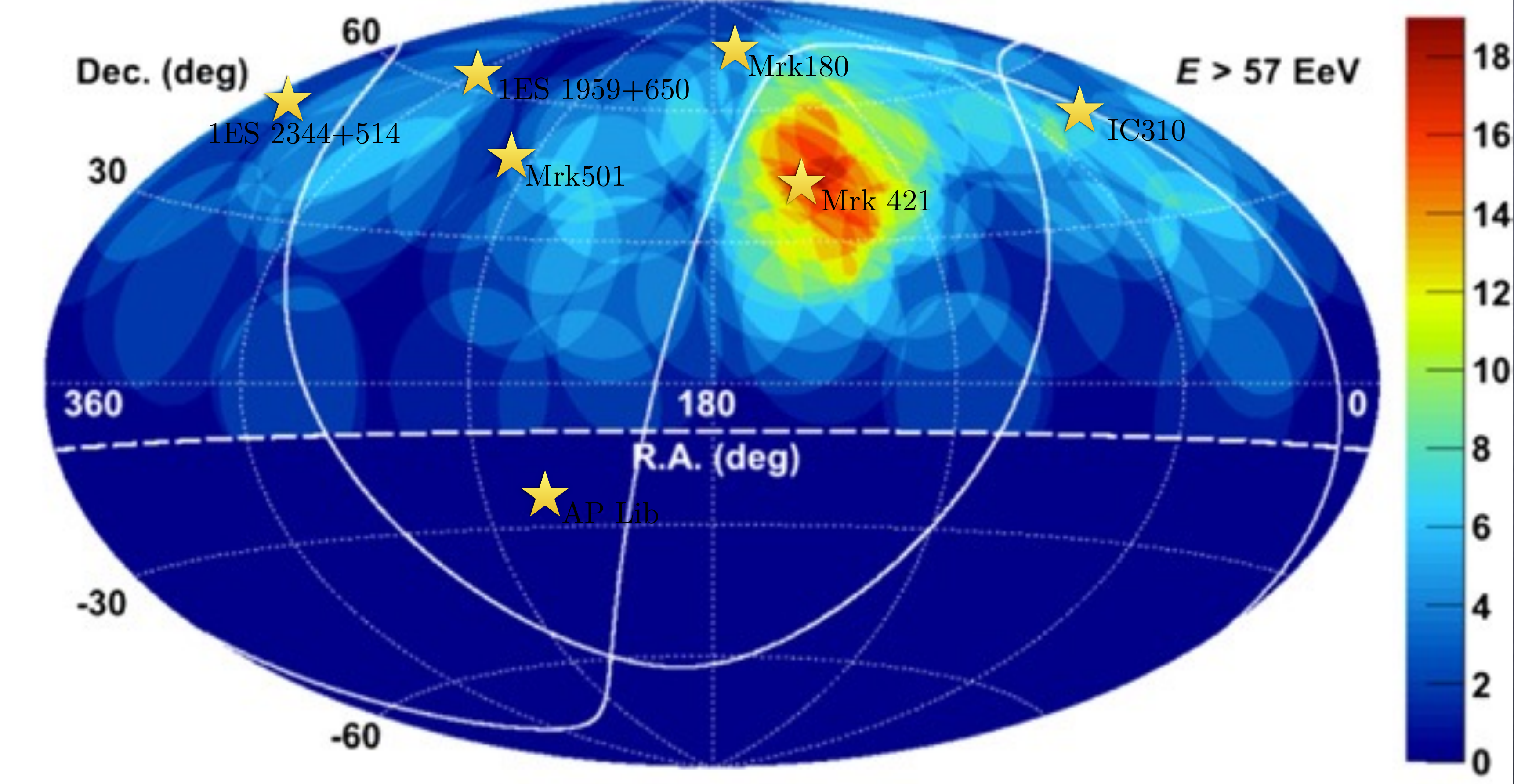}
\caption{Arrival directions of the UHECRs above $5.7\times 10^{19}$eV observed by TA, spread over a $20\deg$ radius circle \citep[From ref.][]{TAhotspot}. Symbols indicate the nearby blazars listed in Table \ref{tab:blazar}. Note the coincidence of the hotspot with the position of Mrk 421.}
\label{fig:TA}
\end{center}
\end{figure}

The directions of arrival of Auger events do not reveal any statistically significant anisotropy \citep{auger13}, while the TA collaboration has found evidence for a hotspot with significance of 3.4$\sigma$ (post-trial) in events above $5.7\times 10^{19}$eV, centered at R.A.=146\fdg 7 and DEC=+43\fdg 2.
Such a quite high level of isotropy may arise just because the Galactic magnetic field of a few $\mu$G could significantly deflect the trajectory of particles with rigidities up to $10^{19}V$ \citep{zirakashvili+98}, i.e., all of the UHECRs if their composition is heavy.
Due to our poor knowledge of intergalactic magnetic fields and incomplete understanding of the magnetic structure of the Milky Way, the details of the propagation of UHECRs are uncertain within orders of magnitude.
We cannot rule out the possibility that the highest-rigidity particles point directly to their sources, in which case the observed quasi-isotropy would imply that the local universe is populated by a relatively large number of sources.

GRBs are transient events and newly-born millisecond pulsars can act as UHECR sources for a very short amount of time, hence it may be extremely hard to backtrack observed UHECR events to such candidates once the delay due to propagation is accounted for. 
On the other hand, for any reasonable model of UHECR propagation, the delay between photon and UHECR arrival times is smaller than the typical AGN duty cycles (several million years), so one can in principle try to correlate \emph{local} active galaxies and UHECR events.

The authors in ref.\ \cite{nemmen+10} have looked for correlations between nearby (within 200\,Mpc, corresponding to redshift $z<0.05$) \emph{GeV-bright} AGNs in the First Fermi-LAT AGN Catalog \cite[1LAC, ][]{1LAC} and early Auger events, reporting a 5.4$\sigma$ correlation on a $\sim 17 \deg$ angular scale. However, it is not clear whether such a correlation improved with the most recent data from Fermi and Auger.
It is interesting to notice that we know only 10 \emph{TeV-bright} AGNs with $z<0.05$ \cite{TeVCat}: Centaurus A, M87, NGC 1275, and the 7 blazars reported in table \ref{tab:blazar}.
Of such 7 blazars, 6 are found in the Northern hemisphere, likely because of the observational strategies of imaging atmospheric Cherenkov telescopes (MAGIC and VERITAS in the North, HESS in the South)\footnote{Milagro reported the detection of Mrk 421 only.}; HAWC and possibly CTA will likely improve the completeness of the census of nearby TeV-bright AGNs, but at this moment there is an asymmetry in the  coverage of the Auger and TA skies in the TeV band. 

Figure \ref{fig:TA} shows the positions of the 7 blazars in table \ref{tab:blazar} on top of the map of TA events above $5.7\times 10^{19}$eV \cite{TAhotspot}.
The most striking feature is that Mrk 421 falls exactly within the TA hotspot (despite the TA discovery paper , while there are no obvious excesses in the directions of the other nearby blazar     s, including the very powerful Mrk 501\footnote{A potential difference between the ``two Markarians'' is that the non-bursting  emission of Mrk 421 is usually modeled by assuming a jet with $\Gamma\sim 30$, a factor of a few larger than Mrk 501's counterpart: this is quite important in the espresso acceleration framework.}. 
We will comment below on the relation between local blazars and the espresso mechanism, but it is interesting to point out that Mrk 421 has a bolometric luminosity of $L_{\rm bol}\approx 2\times 10^{44}\ergs$ and a luminosity distance of about $134$\,Mpc: if its jets generated UHECRs as efficiently as photons, Mrk 421 alone could \emph{in principle} sustain the whole flux of UHECRs detected at Earth ($\approx 3\times 10^{-10}$\ergs\cmq\,sr$^{-1}$ above $10^{18}$eV).
The observed quasi-isotropy of the UHECR flux makes this very unlikely, but such a possibility can be seen as an extreme demonstration of the fact that the unsolved problem of UHECR propagation prevents us from fully exploiting the connections between UHECR anisotropy, flux, and source distribution, especially in the heavy composition scenario.
The correlation between GeV-selected AGNs and Auger data, and the coincidence of Mrk 421 with the marginally significant (3.4$\sigma$ post-trial) TA hotspot do not prove any blazar--UHECRs connection, but it is definitely not inconsistent with the acceleration mechanism proposed in this paper.

\section{Particle Trajectories in Relativistic Flows}\label{sec:traj}
In C15 we studied the orbits of particles gyrating in an infinite relativistic flow with constant magnetic field;
here we generalize such a calculation to the more realistic case of a finite cylindrical flow with an arbitrary toroidal magnetic field.

With a change of notation with respect to C15, throughout the paper we indicate quantities in the laboratory and flow frame respectively with $\tilde{Q}$ and $Q$, and initial/final particle quantities with the subscripts $_{\rm i,f}$; also, we pose $c=1$.
Let us consider a relativistic flow with Lorentz factor $\Gamma$ and velocity $\beta \hat{z}$ with respect to the laboratory frame in cylindrical coordinates ($r$,$\phi$,$z$).
The flow has a radial extent $R$ and a toroidal magnetic field $\textbf{B}= - B(r)\hat{\phi}$, corresponding to a potential vector 
\begin{equation}
\textbf{A}(\textbf{r})= A(r) \hat{z}, \quad\rm{with}\quad  A(r) = - \int_0^{r} B(r') {\rm d} r';
\end{equation}
thus defined, $A(R)= R \langle B\rangle_r$ is related to the radially-averaged toroidal magnetic field.

Consider the Hamiltonian of a particle with mass $m$, charge $q$, and Lorentz factor $\gamma$ in the flow frame:
\begin{equation}
\mathcal{H}= \sqrt{P_r^2 + \frac{P_{\phi}^2}{r^2}+[P_z-qA(r)]^2+m}= \gamma m, 
\end{equation}
where $P_r=\gamma m \dot{r}$, $P_{\phi}=\gamma m r^2 \dot{\phi}$, and $P_z=\gamma m \dot{z} +qA$ are the canonical momenta (the dot indicates the usual time derivative); 
because the Hamiltonian is independent of $t$, $\phi$, and $z$, then $\mathcal{H}$, $P_{\phi}$, and $P_z$ are conserved quantities.
 
We write the momentum of an ultra-relativistic particle with initial energy $\tilde E_{\rm i}$ in the laboratory frame as
\begin{equation}\label{eq:pi}
 \tilde\textbf{p}_{\rm i}\simeq \tilde E_{\rm i}(-\tilde\xii\sqrt{1-\tilde\mui^2}; \sqrt{(1-\tilde\mui^2)(1-\tilde\xii^2)};\tilde\mui),
\end{equation}
where the cosines $\tilde\mui\equiv \tilde p_z/|\tilde\textbf{p}|$ and $\tilde\xii\equiv \tilde p_r/|\tilde\textbf{p}|$ define the ingoing flight angles with respect to the $\hat{z}$ and $-\hat{r}$ axes.

\subsection{Case $\tilde\xii=1$: Planar Orbits}
Let us consider first particles with $\tilde\xii=1$, corresponding to $P_{\phi}=0$ and hence to orbits confined to the $r-z$ plane, which is similar to the case in C15.
The energy gain in the laboratory frame reads (C15, eq.\ 5):
\begin{equation}\label{eq:e}
\mathcal{E}\equiv \frac{\tilde E_{\rm f}}{\tilde E_{\rm i}}= \Gamma^2(1-\beta \tilde\mui)(1+\beta \muf),
\end{equation}
where $\muf\equiv p_{z,{\rm f}}/E$.
From conservation of $P_z$, one has
\begin{equation}
p_z= p_{z,{\rm i}}+q [A(R)-A(r)]
\end{equation}
and therefore (also $E$ being constant):
\begin{equation}
\mathcal{E}= (1-\beta \tilde\mui)
\left\{1+ \Gamma^2 \beta \frac{q}{E}[A(R)-A(r)]\right\}.
\end{equation}
By using $A(R)=R\langle B\rangle_r$ we can introduce the average particle  gyroradius $\mathcal R\equiv E/q\langle B\rangle_r$ and finally write
\begin{equation}
\mathcal{E}= (1-\beta \tilde\mui)
\left\{1+ \Gamma^2 \beta \frac{R}{\mathcal R}\left[1-\frac{A(r)}{A(R)}\right]\right\}.
\end{equation}
The energy gain only depends on $r$, so if a particle comes back to the initial radius $r=R$ its energy gain will be null. 
Here we consider that the particle can be released at any time because of an abrupt change in the flow properties (e.g., it reaches the jet's end), so that the orbit can be truncated at any $r$, when $\mathcal E\geq 1$.
In reality, any perturbation of the perfectly-ordered magnetic field considered here would induce some pitch-angle scattering that leads to a net energy gain.

Since $\mui=(\tilde\mui-\beta)/(1-\tilde\mui\beta)$, the particle will almost always enter the jet moving in the $-\hat z$ direction, and penetrate into the jet for $\sim 2\mathcal R$.
Let us calculate the \emph{maximum} energy gain achievable, $\mathcal E_{\rm max}$, distinguishing the cases $\mathcal R\gtrsim R/2$, in which the particle must cross the entire flow and $\mathcal R\lesssim R/2$, in which the particle ``bounces'' back  to $r=R$. 
In the first case, $\mathcal E_{\rm max}$ is obtained when the particle is on the axis, where $A(r=0)=0$, hence
\begin{equation}\label{eq:elarge}
\mathcal{E}_{\rm max}^{(\mathcal R>R/2)}= (1-\beta \tilde\mui)
\left[1+ \Gamma^2 \beta \frac{R}{\mathcal R}\right]
\simeq
\Gamma^2\frac{R}{\mathcal R}.
\end{equation}
Large-gyroradii particles can only tap a fraction $\frac{R}{\mathcal R}$ of the usual $\sim\Gamma^2$ boost;
the correction $(1-\beta \tilde\mui)$ penalizes particles with momenta in the very direction of the flow, provides an extra factor of 2 for backstreaming particles, and averages to 1 for an isotropic seed distribution.

Particles with $\mathcal R\lesssim R/2$, instead, reach $\mathcal E_{\rm max}$ for a minimum radius $r_{\rm min}$ defined by $P_r=0$, which means that the momentum is entirely along $z$ ($\muf=1$) and so:
\begin{equation}\label{eq:esmall}
\mathcal{E}_{\rm max}^{(\mathcal R<R/2)}= \Gamma^2(1-\beta \tilde\mui)
(1+\beta)\simeq
2\Gamma^2.
\end{equation}
The \emph{average} energy gain along the orbit, $\langle\mathcal E\rangle_t$, obeys $ \mathcal E_{\rm max}\geq\langle\mathcal E\rangle_t\geq \mathcal E_{\rm max}/2$ for any monotonically increasing profile of $B(r)$. 
In fact, $\mathcal E(t)\gtrsim \mathcal E_{\rm max}/2$ for the part of the orbit with $\mu>0$, which encompasses more than half of a gyration ($B$ decreases at small radii); 
$\langle \mathcal E\rangle_t \approx \mathcal E_{\rm max}/2$ in the case of $\mathcal R\ll R$, in which $B\approx$ constant [C15].

\subsection{Case $\tilde\xii<1$: 3D Orbits}
If the particle has a momentum component along $\hat\phi$, the centrifugal barrier due to its non-vanishing angular momentum prevents it from reaching $r=0$, and the orbit has a constant $P_{\phi}$.
$\mathcal E_{\rm max}$  is still achieved when the orbit reaches $\rmin$.
From energy conservation:
\begin{equation}
p_z^2 = \gamma^2\beta^2 m^2-P_{\phi}^2/\rmin^2
\end{equation}
and hence from eq. \ref{eq:e} one gets
\begin{equation}
\mathcal{E}_{\rm max}= \Gamma^2 (1-\beta \tilde\mui)
\left( 1+\beta\sqrt{1-\frac{P_{\phi}^2}{E^2\rmin^2}}\right),
\end{equation}
where, with the definition in eq. \ref{eq:pi} and for $E\simeq\Gamma\tilde E_{\rm i}$,
\begin{equation}
\frac{P_{\phi}^2}{E^2\rmin^2} =
\frac{(1-\tilde\mui^2)(1-\tilde\xii^2)}{\Gamma^2}\frac{R^2}{\rmin^2}\simeq 
\frac{1}{\Gamma^2}\frac{R^2}{\rmin^2}.
\end{equation}
In the limit of $\mathcal R\ll R$ one has $\rmin\gtrsim R-2\mathcal R$, which corresponds to $R^2/\rmin^2\simeq 1+4\mathcal R/R$ and eventually
\begin{equation}\label{eq:phi}
\mathcal{E}_{\rm max}\simeq \Gamma^2 (1-\beta \tilde\mui)
\left\{ 1+\beta\left[1- \frac{1}{\Gamma^2} \left(1+4\frac{\mathcal R}{R}\right)\right]\right\}.
\end{equation}
The correction with respect to eq. \ref{eq:esmall} is small, $\mathcal O(1/\Gamma^2)$, because $p_z$ is boosted in the flow frame, while $p_\phi$ is not.
For the same argument, in the limit of  $\mathcal R\gtrsim R$ the correction to eq.\ \ref{eq:elarge} will be negligible, too.

In summary, for arbitrary toroidal magnetic fields and almost regardless of the ingoing angle, particles that enter the flow with energy $\tilde E_{\rm i}$  are accelerated by a factor of $\gtrsim\Gamma^2$ for most of their orbit if their average gyroradius in the flow frame $\mathcal R\simeq\Gamma \tilde E_{\rm i}/q\langle B\rangle_r\lesssim R/2$, or up to $\sim\Gamma^2R/\mathcal R$ for larger gyroradii.
Since $\tilde \mathcal R\simeq E_{\rm PeV}/B_{\mu{\rm G}}$\,pc the full energy gain is expected for all of the Galactic CRs if $R\gtrsim \Gamma/B_{\mu{\rm G}}$\,pc, which is not a tight constraint.

If the particle is not released during the first orbit, it will reach $r=R$ after having travelled a distance $\Delta z\simeq\Gamma T\simeq\Gamma^2 \tilde \mathcal R$, where $T\simeq 2\pi \mathcal R$ is the Larmor period in the flow frame.
Knee CRs  may travel distances as large as a few kpc, comparable with the jet's extent, in a single orbit; 
lower-energy CRs, instead, will likely enter and exit the jet several times before being released.

\subsection{The General Case}
\begin{figure}[t]
\begin{center}
\includegraphics[trim=80px 70px 80px 90px, clip=true, width=.5\textwidth]{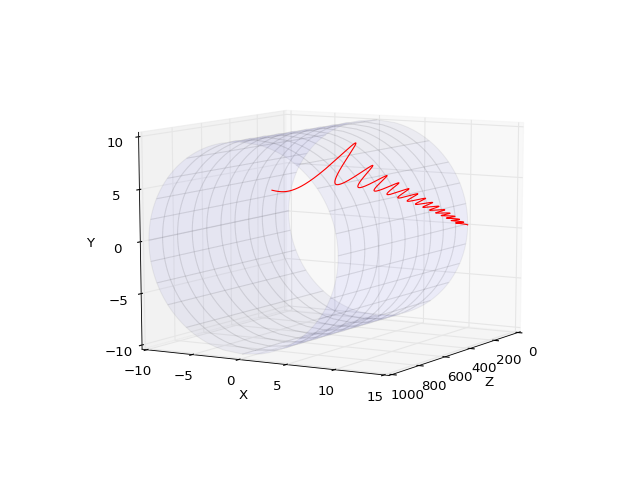}
\includegraphics[trim=71px 0px 65px 0px, clip=true, width=.5\textwidth]{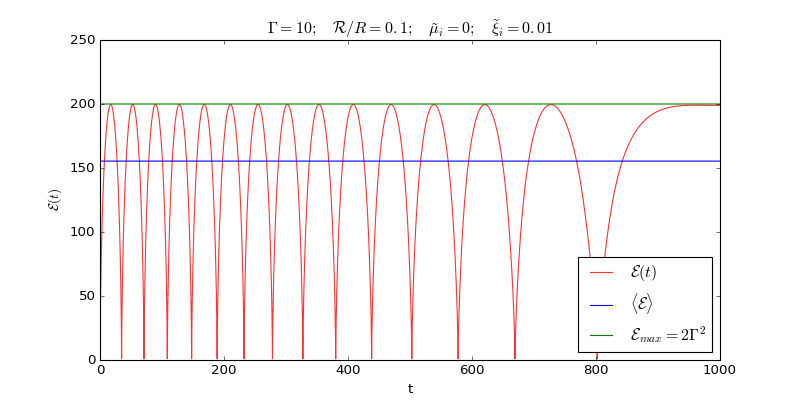}
\caption{Top panel: Example of the trajectory of a particle with initial $\tilde\gamma_{\rm i}=50$ injected into a relativistic jet with $\Gamma=10$ (see text for more details).
Bottom panel: energy gain $\mathcal E$ in the laboratory frame during the particle's orbit. Note that the average energy gain is $\sim 1.5\Gamma^2$.}
\label{fig:part}
\end{center}
\end{figure}

We consider now a cylindrical jet with $\Gamma=10$ and length $Z=100R$, with a toroidal magnetic field
\begin{equation}\label{eq:B}
{\bf B}(r\leq R,z)= B_0\frac{z-Z}{Z}\frac{r}{R}\hat\phi.
\end{equation}
In the laboratory frame, in addition to the boosted field, we add a decreasing magnetic field $\tilde B(r\geq R,z)=B(R,z)R/r$ to ensure continuity and allow particles to get in and out of the jet multiple times. 
A particle with initial $\tilde\gamma_{\rm i}=50$ (corresponding to $\mathcal R=0.1R$) is injected into the jet at $(r,z)=(R,0)$ with $\tilde\mui=0$ and finite angular momentum due to a non vanishing $\tilde p_\phi$ ($\tilde\xii=0.01$, see  eq.\ \ref{eq:pi}).
Its trajectory and energy gain, shown in fig.\ \ref{fig:part}, are just representative of many possible ones, but for the arguments above we expect almost all of the orbits to lead to similar energy gains.
 
We see that, while the particle is ``embroidering'' the jet surface,  its energy oscillates between $\tilde E_{\rm i}$ and $2\Gamma^2\tilde E_{\rm i}$, as predicted by eq.\ \ref{eq:esmall}, with an average gain of $\langle \mathcal E\rangle_t \approx  \pi/2 \Gamma^2$ determined by  the chosen profile $B(r\leq R)\propto r$.
In this example the gyroradius of the particle becomes larger and larger, until it is released at the jet's end.
The effect of the decreasing field is to induce a radial $\nabla B$-drift that does not affect the orbit significantly.
Adding a poloidal magnetic field or a different scaling with $r,z$ does not change our main results appreciably.

In reality, relativistic jets are expected to be stratified, their structure depending on the stabilization mechanism \cite[e.g.,][]{mignone+10,pk15,tb16};
therefore, particles with small $\mathcal R$ may initially probe only relatively-small Lorentz factors.
Then, thanks to magnetic/velocity inhomogeneities, such particles may be energized at any flow crossing, probing deeper and deeper layers while propagating along the jet. 
The final spectrum of espresso-accelerated particles will hence depend on the jet structure and will be addressed in a forthcoming work \cite{jet}.

\section{A Few Remarks}
In C15 we discussed the espresso mechanism in relation to other models in the literature, as well as its applicability to environments other than AGN jets; here we outline two additional points not  covered previously. 

In the calculations above we set $B_{\phi}<0$, which is required for a charged nucleus to be accelerated toward the jet's end, while in reality the sign of $B_{\phi}$ is determined by the current in the jet, which depends on the jet nature (whether it is kinetically/magnetically and hadron/pair dominated).   
Acceleration with a $\sim\Gamma^2$ boost is in principle possible also if nuclei were shot toward the black hole, even if the intense radiation field close to the AGN may be detrimental for UHECR survival. 
Note that also \emph{CR electrons} can be accelerated via the same process, being shot in the opposite direction. 
Such electrons will suffer prominent synchrotron and inverse-Compton losses: they may not be able to tap the full $\sim\Gamma^2$ boost (or in general to reach multi-TeV energies), but the radiation that they emit in the process may sizably contribute to the AGN luminosity.
We will present more quantitative calculations in a future work, but it is worth stressing that the proposed mechanism may represent a very ubiquitous energization process that does not require either shocks or magnetic reconnection. 

Concerning the discussion in \S\ref{sec:sources}, it is possible that most of the UHECRs that hit the Earth may not come from AGNs that look like  \emph{blazars}: typical jets observed in FR-I and FR-II radio galaxies terminate in several-kpc-wide lobes, with inferred magnetic fields of tens of $\mu$G, which are sufficient for isotropizing the relativistically-beamed UHECR distribution expected at the end of ultra-relativistic jet.
Therefore, while we might still expect UHECR hotspots coincident with local blazars, a contribution to the quasi-isotropic UHECR flux may  also come from nearby non-blazar AGNs with high-$\Gamma$ jets not pointing directly at us. 

In summary, we outlined the properties of the re-acceleration of ordinary CRs produced in SNRs in ultra-relativistic AGN jets, suggesting that this so-called \emph{espresso} mechanism may generate UHECRs.
Such a model naturally accounts for the UHECR chemical composition measured by Auger and is consistent with the distribution of AGNs in the local universe, especially with the TA hotspot in events with $E\gtrsim 5.7\times 10^{19}$\,eV, which coincides with the position of Mrk 421.
We have also presented analytical and numerical trajectories of re-accelerated particles in  simplified jet structures, confirming that almost all of the particles that make it to the high-$\Gamma$ region of the jet get an average energy boost $\gtrsim \Gamma^2$.

\vspace{.5cm}
The author wishes to thank the organizers and the participants of the San Vito CRBTSM 2016 Conference for the outstanding venue and the lively discussions, and Rostom Mbarek (U.\ of Chicago) for producing the Python scripts that lead to the plots in fig.\ \ref{fig:part}.




\bibliographystyle{elsarticle-num}
\bibliography{UHECRs}

\begin{thebibliography}{10}
\expandafter\ifx\csname url\endcsname\relax
  \def\url#1{\texttt{#1}}\fi
\expandafter\ifx\csname urlprefix\endcsname\relax\def\urlprefix{URL }\fi
\expandafter\ifx\csname href\endcsname\relax
  \def\href#1#2{#2} \def\path#1{#1}\fi

\bibitem{tycho}
G.~{Morlino}, D.~{Caprioli}, {Strong evidence for hadron acceleration in
  Tycho's supernova remnant}, A\&A 538 (2012) A81.
\newblock \href {http://arxiv.org/abs/arXiv:1105.6342}
  {\path{arXiv:arXiv:1105.6342}}, \href
  {http://dx.doi.org/10.1051/0004-6361/201117855}
  {\path{doi:10.1051/0004-6361/201117855}}.

\bibitem{DSA}
D.~{Caprioli}, A.~{Spitkovsky}, {Simulations of Ion Acceleration at
  Non-relativistic Shocks: I. Acceleration Efficiency}, \apj 783 (2014) 91.
\newblock \href {http://arxiv.org/abs/1310.2943} {\path{arXiv:1310.2943}},
  \href {http://dx.doi.org/10.1088/0004-637X/783/2/91}
  {\path{doi:10.1088/0004-637X/783/2/91}}.

\bibitem{MFA}
D.~{Caprioli}, A.~{Spitkovsky}, {Simulations of Ion Acceleration at
  Non-relativistic Shocks: II. Magnetic Field Amplification}, \apj 794 (2014)
  46.
\newblock \href {http://arxiv.org/abs/1401.7679} {\path{arXiv:1401.7679}},
  \href {http://dx.doi.org/10.1088/0004-637X/794/1/46}
  {\path{doi:10.1088/0004-637X/794/1/46}}.

\bibitem{diffusion}
D.~{Caprioli}, A.~{Spitkovsky}, {Simulations of Ion Acceleration at
  Non-relativistic Shocks. III. Particle Diffusion}, \apj 794 (2014) 47.
\newblock \href {http://arxiv.org/abs/1407.2261} {\path{arXiv:1407.2261}},
  \href {http://dx.doi.org/10.1088/0004-637X/794/1/47}
  {\path{doi:10.1088/0004-637X/794/1/47}}.

\bibitem{injection}
D.~{Caprioli}, A.~{Pop}, A.~{Spitkovsky}, {Simulations and Theory of Ion
  Injection at Non-relativistic Collisionless Shocks}, \apjl 798 (2015) 28.
\newblock \href {http://arxiv.org/abs/1409.8291} {\path{arXiv:1409.8291}}.

\bibitem{AZ}
D.~Caprioli, D.~T. Yi, A.~Spitkovsky, Chemical enhancements in
  shock-accelerated particles: Ab-initio simulations, ArXiv e-prints\href
  {http://arxiv.org/abs/1704.08252} {\path{arXiv:1704.08252}}.

\bibitem{hoerandel+06}
J.~R. {H{\"o}randel et al.}, {Results from the KASCADE, KASCADE-Grande, and
  LOPES experiments}, Journal of Physics Conference Series 39 (2006) 463--470.
\newblock \href {http://arxiv.org/abs/astro-ph/0511649}
  {\path{arXiv:astro-ph/0511649}}, \href
  {http://dx.doi.org/10.1088/1742-6596/39/1/122}
  {\path{doi:10.1088/1742-6596/39/1/122}}.

\bibitem{argo12}
B.~{Bartoli et al. [ARGO-YBJ Collaboration]}, {Light-component spectrum of the
  primary cosmic rays in the multi-TeV region measured by the ARGO-YBJ
  experiment}, \prd 85~(9) (2012) 092005.
\newblock \href {http://dx.doi.org/10.1103/PhysRevD.85.092005}
  {\path{doi:10.1103/PhysRevD.85.092005}}.

\bibitem{nuclei}
D.~{Caprioli}, P.~{Blasi}, E.~{Amato}, {Non-linear diffusive acceleration of
  heavy nuclei in supernova remnant shocks}, APh 34 (2011) 447--456.
\newblock \href {http://arxiv.org/abs/1007.1925} {\path{arXiv:1007.1925}},
  \href {http://dx.doi.org/10.1016/j.astropartphys.2010.10.011}
  {\path{doi:10.1016/j.astropartphys.2010.10.011}}.

\bibitem{waxman95}
E.~{Waxman}, {Cosmological Gamma-Ray Bursts and the Highest Energy Cosmic
  Rays}, Physical Review Letters 75 (1995) 386--389.
\newblock \href {http://arxiv.org/abs/astro-ph/9505082}
  {\path{arXiv:astro-ph/9505082}}, \href
  {http://dx.doi.org/10.1103/PhysRevLett.75.386}
  {\path{doi:10.1103/PhysRevLett.75.386}}.

\bibitem{vietri95}
M.~{Vietri}, {The Acceleration of Ultra--High-Energy Cosmic Rays in Gamma-Ray
  Bursts}, \apj 453 (1995) 883.
\newblock \href {http://arxiv.org/abs/astro-ph/9506081}
  {\path{arXiv:astro-ph/9506081}}, \href {http://dx.doi.org/10.1086/176448}
  {\path{doi:10.1086/176448}}.

\bibitem{Blasi+00}
P.~{Blasi}, R.~I. {Epstein}, A.~V. {Olinto}, {Ultra-High-Energy Cosmic Rays
  from Young Neutron Star Winds}, ApJ Lett. 533 (2000) L123--L126.
\newblock \href {http://arxiv.org/abs/astro-ph/9912240}
  {\path{arXiv:astro-ph/9912240}}, \href {http://dx.doi.org/10.1086/312626}
  {\path{doi:10.1086/312626}}.

\bibitem{Fang+12}
K.~{Fang}, K.~{Kotera}, A.~V. {Olinto}, {Newly Born Pulsars as Sources of
  Ultrahigh Energy Cosmic Rays}, \apj 750 (2012) 118.
\newblock \href {http://arxiv.org/abs/1201.5197} {\path{arXiv:1201.5197}},
  \href {http://dx.doi.org/10.1088/0004-637X/750/2/118}
  {\path{doi:10.1088/0004-637X/750/2/118}}.

\bibitem{ostrowsky00}
M.~{Ostrowski}, {On possible `cosmic ray cocoons' of relativistic jets}, MNRAS
  312 (2000) 579--584.
\newblock \href {http://arxiv.org/abs/astro-ph/9910491}
  {\path{arXiv:astro-ph/9910491}}, \href
  {http://dx.doi.org/10.1046/j.1365-8711.2000.03146.x}
  {\path{doi:10.1046/j.1365-8711.2000.03146.x}}.

\bibitem{Murase+12}
K.~{Murase}, C.~D. {Dermer}, H.~{Takami}, G.~{Migliori}, {Blazars as
  Ultra-high-energy Cosmic-ray Sources: Implications for TeV Gamma-Ray
  Observations}, \apj 749 (2012) 63.
\newblock \href {http://arxiv.org/abs/1107.5576} {\path{arXiv:1107.5576}},
  \href {http://dx.doi.org/10.1088/0004-637X/749/1/63}
  {\path{doi:10.1088/0004-637X/749/1/63}}.

\bibitem{espresso}
D.~{Caprioli}, {''Espresso'' Acceleration of Ultra-high-energy Cosmic Rays},
  \apjl 811 (2015) L38.
\newblock \href {http://arxiv.org/abs/1505.06739} {\path{arXiv:1505.06739}},
  \href {http://dx.doi.org/10.1088/2041-8205/811/2/L38}
  {\path{doi:10.1088/2041-8205/811/2/L38}}.

\bibitem{Auger14a}
A.~{Aab et al. [Auger Collaboration]}, {Depth of maximum of air-shower profiles
  at the Pierre Auger Observatory. I. Measurements at energies above
  10$^{17.8}$ eV}, \prd 90~(12) (2014) 122005.
\newblock \href {http://dx.doi.org/10.1103/PhysRevD.90.122005}
  {\path{doi:10.1103/PhysRevD.90.122005}}.

\bibitem{Auger16}
A.~{Aab et al. [Auger Collaboration]}, {Evidence for a mixed mass composition
  at the 'ankle' in the cosmic-ray spectrum}, Physics Letters B 762 (2016)
  288--295.
\newblock \href {http://arxiv.org/abs/1609.08567} {\path{arXiv:1609.08567}},
  \href {http://dx.doi.org/10.1016/j.physletb.2016.09.039}
  {\path{doi:10.1016/j.physletb.2016.09.039}}.

\bibitem{Pierog13}
T.~{Pierog}, {LHC results and High Energy Cosmic Ray Interaction Models},
  Journal of Physics Conference Series 409~(1) (2013) 012008.
\newblock \href {http://dx.doi.org/10.1088/1742-6596/409/1/012008}
  {\path{doi:10.1088/1742-6596/409/1/012008}}.

\bibitem{PAO-TA15}
R.~{Abbasi et al. [Auger and Telescope Array Collaborations]}, {Report of the
  Working Group on the Composition of Ultra High Energy Cosmic Rays}, ArXiv
  e-prints\href {http://arxiv.org/abs/1503.07540} {\path{arXiv:1503.07540}}.

\bibitem{Achterberg+01}
A.~{Achterberg}, Y.~A. {Gallant}, J.~G. {Kirk}, A.~W. {Guthmann}, {Particle
  acceleration by ultrarelativistic shocks: theory and simulations}, MNRAS 328
  (2001) 393--408.
\newblock \href {http://arxiv.org/abs/astro-ph/0107530}
  {\path{arXiv:astro-ph/0107530}}, \href
  {http://dx.doi.org/10.1046/j.1365-8711.2001.04851.x}
  {\path{doi:10.1046/j.1365-8711.2001.04851.x}}.

\bibitem{Tavecchio+10}
F.~{Tavecchio}, G.~{Ghisellini}, G.~{Ghirlanda}, L.~{Foschini}, L.~{Maraschi},
  {TeV BL Lac objects at the dawn of the Fermi era}, MNRAS 401 (2010)
  1570--1586.
\newblock \href {http://arxiv.org/abs/0909.0651} {\path{arXiv:0909.0651}},
  \href {http://dx.doi.org/10.1111/j.1365-2966.2009.15784.x}
  {\path{doi:10.1111/j.1365-2966.2009.15784.x}}.

\bibitem{KU12}
K.-H. {Kampert}, M.~{Unger}, {Measurements of the cosmic ray composition with
  air shower experiments}, \APh 35 (2012) 660--678.
\newblock \href {http://arxiv.org/abs/1201.0018} {\path{arXiv:1201.0018}},
  \href {http://dx.doi.org/10.1016/j.astropartphys.2012.02.004}
  {\path{doi:10.1016/j.astropartphys.2012.02.004}}.

\bibitem{GST13}
T.~K. {Gaisser}, T.~{Stanev}, S.~{Tilav}, {Cosmic ray energy spectrum from
  measurements of air showers}, Frontiers of Physics 8 (2013) 748--758.
\newblock \href {http://arxiv.org/abs/1303.3565} {\path{arXiv:1303.3565}},
  \href {http://dx.doi.org/10.1007/s11467-013-0319-7}
  {\path{doi:10.1007/s11467-013-0319-7}}.

\bibitem{Auger14b}
A.~{Aab et al. [Auger Collaboration]}, {Depth of maximum of air-shower profiles
  at the Pierre Auger Observatory. II. Composition implications}, \prd 90~(12)
  (2014) 122006.
\newblock \href {http://dx.doi.org/10.1103/PhysRevD.90.122006}
  {\path{doi:10.1103/PhysRevD.90.122006}}.

\bibitem{abb14}
R.~{Aloisio}, V.~{Berezinsky}, P.~{Blasi}, {Ultra high energy cosmic rays:
  implications of Auger data for source spectra and chemical composition},
  \jcap 10 (2014) 20.
\newblock \href {http://arxiv.org/abs/1312.7459} {\path{arXiv:1312.7459}},
  \href {http://dx.doi.org/10.1088/1475-7516/2014/10/020}
  {\path{doi:10.1088/1475-7516/2014/10/020}}.

\bibitem{Taylor14}
A.~M. {Taylor}, {UHECR composition models}, \APh 54 (2014) 48--53.
\newblock \href {http://arxiv.org/abs/1401.0199} {\path{arXiv:1401.0199}},
  \href {http://dx.doi.org/10.1016/j.astropartphys.2013.11.006}
  {\path{doi:10.1016/j.astropartphys.2013.11.006}}.

\bibitem{mr13}
S.~{Mollerach}, E.~{Roulet}, {Magnetic diffusion effects on the ultra-high
  energy cosmic ray spectrum and composition}, \jcap 10 (2013) 013.
\newblock \href {http://arxiv.org/abs/1305.6519} {\path{arXiv:1305.6519}},
  \href {http://dx.doi.org/10.1088/1475-7516/2013/10/013}
  {\path{doi:10.1088/1475-7516/2013/10/013}}.

\bibitem{as14}
R.~{Alves Batista}, G.~{Sigl}, {Diffusion of cosmic rays at EeV energies in
  inhomogeneous extragalactic magnetic fields}, \jcap 11 (2014) 031.
\newblock \href {http://arxiv.org/abs/1407.6150} {\path{arXiv:1407.6150}},
  \href {http://dx.doi.org/10.1088/1475-7516/2014/11/031}
  {\path{doi:10.1088/1475-7516/2014/11/031}}.

\bibitem{murase09}
K.~{Murase}, Ultrahigh-energy photons as a probe of nearby transient
  ultrahigh-energy cosmic-ray sources and possible lorentz-invariance
  violation, \prl 103 (2009) 081102.
\newblock \href {http://dx.doi.org/10.1103/PhysRevLett.103.081102}
  {\path{doi:10.1103/PhysRevLett.103.081102}}.

\bibitem{Katz+13}
B.~{Katz}, E.~{Waxman}, T.~{Thompson}, A.~{Loeb}, {The energy production rate
  density of cosmic rays in the local universe is
  \$$\backslash$sim10\^{}$\{$44-45$\}$$\backslash$rm
  erg\~{}Mpc\^{}$\{$-3$\}$\~{}yr\^{}$\{$-1$\}$\$ at all particle energies},
  ArXiv e-prints\href {http://arxiv.org/abs/1311.0287}
  {\path{arXiv:1311.0287}}.

\bibitem{Jiang+07}
L.~{Jiang}, X.~{Fan}, {\v Z}.~{Ivezi{\'c}}, G.~T. {Richards}, D.~P.
  {Schneider}, M.~A. {Strauss}, B.~C. {Kelly}, {The Radio-Loud Fraction of
  Quasars is a Strong Function of Redshift and Optical Luminosity}, \apj 656
  (2007) 680--690.
\newblock \href {http://arxiv.org/abs/astro-ph/0611453}
  {\path{arXiv:astro-ph/0611453}}, \href {http://dx.doi.org/10.1086/510831}
  {\path{doi:10.1086/510831}}.

\bibitem{woo-urry02}
J.-H. {Woo}, C.~M. {Urry}, {Active Galactic Nucleus Black Hole Masses and
  Bolometric Luminosities}, \apj 579 (2002) 530--544.
\newblock \href {http://arxiv.org/abs/astro-ph/0207249}
  {\path{arXiv:astro-ph/0207249}}, \href {http://dx.doi.org/10.1086/342878}
  {\path{doi:10.1086/342878}}.

\bibitem{GTG09}
G.~{Ghisellini}, F.~{Tavecchio}, G.~{Ghirlanda}, {Jet and accretion power in
  the most powerful Fermi blazars}, MNRAS 399 (2009) 2041--2054.
\newblock \href {http://arxiv.org/abs/0906.2195} {\path{arXiv:0906.2195}},
  \href {http://dx.doi.org/10.1111/j.1365-2966.2009.15397.x}
  {\path{doi:10.1111/j.1365-2966.2009.15397.x}}.

\bibitem{Waxman04}
E.~{Waxman}, {Extra-galactic sources of high-energy neutrinos}, New Journal of
  Physics 6 (2004) 140.
\newblock \href {http://dx.doi.org/10.1088/1367-2630/6/1/140}
  {\path{doi:10.1088/1367-2630/6/1/140}}.

\bibitem{hillas84}
A.~M. {Hillas}, {The Origin of Ultra-High-Energy Cosmic Rays}, \araa 22 (1984)
  425--444.
\newblock \href {http://dx.doi.org/10.1146/annurev.aa.22.090184.002233}
  {\path{doi:10.1146/annurev.aa.22.090184.002233}}.

\bibitem{greisen66}
K.~{Greisen}, {End to the Cosmic-Ray Spectrum?}, Physical Review Letters 16
  (1966) 748--750.
\newblock \href {http://dx.doi.org/10.1103/PhysRevLett.16.748}
  {\path{doi:10.1103/PhysRevLett.16.748}}.

\bibitem{ZK66}
G.~T. {Zatsepin}, V.~A. {Kuz'min}, {Upper Limit of the Spectrum of Cosmic
  Rays}, ZhETF Pis'ma Redaktsiiu 4 (1966) 114--+.

\bibitem{Dermer07}
C.~D. {Dermer}, {On Gamma Ray Burst and Blazar AGN Origins of the Ultra-High
  Energy Cosmic Rays in Light of First Results from Auger}, ArXiv e-prints\href
  {http://arxiv.org/abs/0711.2804} {\path{arXiv:0711.2804}}.

\bibitem{ko11}
K.~{Kotera}, A.~V. {Olinto}, {The Astrophysics of Ultrahigh-Energy Cosmic
  Rays}, \araa 49 (2011) 119--153.
\newblock \href {http://arxiv.org/abs/1101.4256} {\path{arXiv:1101.4256}},
  \href {http://dx.doi.org/10.1146/annurev-astro-081710-102620}
  {\path{doi:10.1146/annurev-astro-081710-102620}}.

\bibitem{TAhotspot}
R.~U. {Abbasi et al.}, {Indications of Intermediate-scale Anisotropy of Cosmic
  Rays with Energy Greater Than 57 EeV in the Northern Sky Measured with the
  Surface Detector of the Telescope Array Experiment}, ApJ Lett. 790 (2014)
  L21.
\newblock \href {http://dx.doi.org/10.1088/2041-8205/790/2/L21}
  {\path{doi:10.1088/2041-8205/790/2/L21}}.

\bibitem{auger13}
P.~{Abreu et al. [Auger Collaboration]}, {Constraints on the Origin of Cosmic
  Rays above 10$^{18}$ eV from Large-scale Anisotropy Searches in Data of the
  Pierre Auger Observatory}, \apjl 762 (2013) L13.
\newblock \href {http://arxiv.org/abs/1212.3083} {\path{arXiv:1212.3083}},
  \href {http://dx.doi.org/10.1088/2041-8205/762/1/L13}
  {\path{doi:10.1088/2041-8205/762/1/L13}}.

\bibitem{zirakashvili+98}
V.~N. {Zirakashvili}, D.~N. {Pochepkin}, V.~S. {Ptuskin}, S.~I. {Rogovaya},
  {Propagation of ultra-high-energy cosmic rays in Galactic magnetic fields},
  Astronomy Letters 24 (1998) 139--143.

\bibitem{nemmen+10}
R.~S. {Nemmen}, C.~{Bonatto}, T.~{Storchi-Bergmann}, {A Correlation Between the
  Highest Energy Cosmic Rays and Nearby Active Galactic Nuclei Detected by
  Fermi}, \apj 722 (2010) 281--288.
\newblock \href {http://arxiv.org/abs/1007.5317} {\path{arXiv:1007.5317}},
  \href {http://dx.doi.org/10.1088/0004-637X/722/1/281}
  {\path{doi:10.1088/0004-637X/722/1/281}}.

\bibitem{1LAC}
A.~A. {Abdo et al. [Fermi Collaboration]}, {The First Catalog of Active
  Galactic Nuclei Detected by the Fermi Large Area Telescope}, \apj 715 (2010)
  429--457.
\newblock \href {http://arxiv.org/abs/1002.0150} {\path{arXiv:1002.0150}},
  \href {http://dx.doi.org/10.1088/0004-637X/715/1/429}
  {\path{doi:10.1088/0004-637X/715/1/429}}.

\bibitem{TeVCat}
S.~P. {Wakely}, D.~{Horan}, {TeVCat: An online catalog for Very High Energy
  Gamma-Ray Astronomy}, International Cosmic Ray Conference 3 (2008)
  1341--1344.

\bibitem{mignone+10}
A.~{Mignone}, P.~{Rossi}, G.~{Bodo}, A.~{Ferrari}, S.~{Massaglia},
  {High-resolution 3D relativistic MHD simulations of jets}, \mnras 402 (2010)
  7--12.
\newblock \href {http://arxiv.org/abs/0908.4523} {\path{arXiv:0908.4523}},
  \href {http://dx.doi.org/10.1111/j.1365-2966.2009.15642.x}
  {\path{doi:10.1111/j.1365-2966.2009.15642.x}}.

\bibitem{pk15}
O.~{Porth}, S.~S. {Komissarov}, {Causality and stability of cosmic jets},
  \mnras 452 (2015) 1089--1104.
\newblock \href {http://arxiv.org/abs/1408.3318} {\path{arXiv:1408.3318}},
  \href {http://dx.doi.org/10.1093/mnras/stv1295}
  {\path{doi:10.1093/mnras/stv1295}}.

\bibitem{tb16}
A.~{Tchekhovskoy}, O.~{Bromberg}, {Three-dimensional relativistic MHD
  simulations of active galactic nuclei jets: magnetic kink instability and
  Fanaroff-Riley dichotomy}, \mnras 461 (2016) L46--L50.
\newblock \href {http://arxiv.org/abs/1512.04526} {\path{arXiv:1512.04526}},
  \href {http://dx.doi.org/10.1093/mnrasl/slw064}
  {\path{doi:10.1093/mnrasl/slw064}}.

\bibitem{jet}
D.~{Caprioli}, R.~{Mbarek}, {\emph{Espresso} acceleration in 3D MHD simulations
  of relativistic jets}, in prog.

\end{thebibliography}







\end{document}